**Nanopore creation in graphene at the nanoscale for water desalination.**

Sidi Abdelmajid Ait Abdelkader[1,+], Ismail Benabdallah[1,+], Mohammed Amlieh[1,+], Abdelouahad El Fatimy[1]

[1] Institute of Applied Physics, Mohammed VI Polytechnic University, Lot 660, Hay Moulay Rachid, Ben Guerir, 43150, Morocco.

[+] These authors contributed equally

[*] Corresponding author: Abdelouahed.ELFATIMY@um6p.ma

Creating nanopores in graphene is a powerful tool for engineering its properties. Nanopores in graphene tune their electrical, optical, magnetic, and mechanical properties. However, controlling nanopores formation at the nanoscale level remains a significant challenge. We report an easy method to control nanopore sizes using argon-plasma magnetron sputtering. By calculating and measuring Raman spectra, we show that the nano-pores in graphene are controllable and size-tunable. Furthermore, we report that the graphene Raman mode around 1450 cm$^{-1}$, which was attributed to the substrate effect, is due to nanopores.

Here, we also propose a novel graphene device-based water filtration. Our proposed concept of two graphene electrodes with nanopores on the substrate (SiC and SiO$_2$) makes it possible to have the highest permeability value, keeping almost a 100 % salt rejection and improving its mechanical properties. These reported results are essential for developing water desalination membranes based on graphene devices.

The 2018 United Nations water development reported that around six billion people will suffer from clean water scarcity by 2050. This result is due to increased water demand with water resource reduction and pollution driven by dramatic population and economic growth[1]. Therefore, desalinating seawater is urgently needed, as it represents 97.5 % of the planet's total water.[2] One of the most promising and energy-efficient technologies for desalination is reverse osmosis (RO)[2-3]. In the RO process, an external hydrostatic pressure forces water through a porous membrane. The permeable membrane is the critical element of the RO.

**Table 1. Classification of recent studies on Na$^+$ and Cl$^-$ filtration through nanoporous monolayer graphene for the hydrogenated pore. (n.a. means not applicable).**

| Membrane | Pore diameter | Permeability (L/m². h. bar) | Salt Rejection | Pressure (MPa) | Electric Field (V/Å) | Ref. |
|---|---|---|---|---|---|---|
| Graphene | 0.87 nm | 2500 | 70% | 150 | n.a | [4] |
| Graphene | 1.46 nm | 3630 | 60% | 100 | n.a | [6] |
| Graphene | 0.82 nm | 785.6 | 93% | 10-200 | n.a | [7] |
| Graphene/Si | ~ 1 nm | 2.6 ml/day | 100% | 5 | n.a | [8] |
| Two-electrode graphene/SiO$_2$ | 0.81 nm / 1.0 nm* | 894 / 1500* | 100% / 93.75%* | 50 – 1000 | 0 / 2.5 | This work |
| Two-electrode Graphene/SiC | 0.81 nm / 1.0 nm* | 1208 / 2333* | 100% / 100%* | 50 – 1000 | 0 / 2.5 | This work |
| Two-electrode Graphene | 0.81 nm / 1.0 nm* | 2015 / 2708* | 100% / 93.75%* | 50 – 1000 | 0 / 2.5 | This work |

*For the estimation methods, see supplementary information.

2D Materials based graphene has an excellent potential to be used as a RO membrane due to its unique properties, such as mechanical and chemical stability, and its atomic thickness, which allows it to transport water faster than any conventional RO membrane, with a water flux scaling inversely to membrane thickness (Table.1).[2,3] Tanugi et al. reported that nanometer-scale pores in single-layer freestanding graphene could effectively filter NaCl Salt from water; the water permeability reported is about 2500 L/m².h.bar for the hydrogenated pore and 4167 L/m².h.bar for the hydroxylated pore with a salt rejection of about 70% and 45%, respectively at 150 MPa applied pressure by molecular dynamics simulation study. These values are several orders of magnitude higher than conventional reverse osmosis membranes.[4-5] Table 1 lists the central report concept works using a similar idea of Tanugi et al. with a single-layer freestanding mono-layer graphene membrane with different pore geometries and for the hydrogenated pore only for comparison. However, freestanding mono-layer graphene is not mechanically stable at high pressure and does not meet the industrial requirements and scale-up. Therefore, a more realistic approach and easy fabrication remain a big challenge.

Here, we propose a novel water filtration membrane based on two-electrode graphene with nanopores on substrates. We studied the effect of substrates on the permeability and salt rejection values for the two-electrode graphene membrane with nanopore-based on $SiO_2$ (2E-NPG /$SiO_2$) and SiC (2E-NPG/SiC) substrates with and without an applied electric field. Using Molecular Dynamics, we report a 100 % salt rejection with a maximum water permeability of about 1208 and 2333 for the 2E-NPG/SiC for the hydrogenated pore size of about 0.8 nm and about 1 nm. The desalination performance can be higher with a larger nanopore thanks to the electric field effect on the salt rejection rate. Furthermore, we propose an easy way to create nanopores at the Nanoscale level that are produced with high yield when a protective metal layer is deposited directly on the as-grown graphene or chemical vapor deposition graphene. This layer also prevents graphene contact with organic residues. We show that this method allows the creation and control of nanopores at the nanoscale. We show that this sputtering process can create nano-pore in graphene with sizes from 0.75 nm to 1.25 nm and more significant if we add more power to sputtering. Moreover, we find that the Raman mode (around 1450 cm$^{-1}$) reported and attributed in the literature to the substrate effect due to graphene defects.

We sputtered 30 nm of gold using argon-plasma magnetron sputtering with different powers, as shown in Figure 1 (a). Then, the deposited layer of metal was subsequently removed by aqua regia[9-10]. This allows a one-step membrane fabrication process by protecting the two electrodes from aqua Regia.

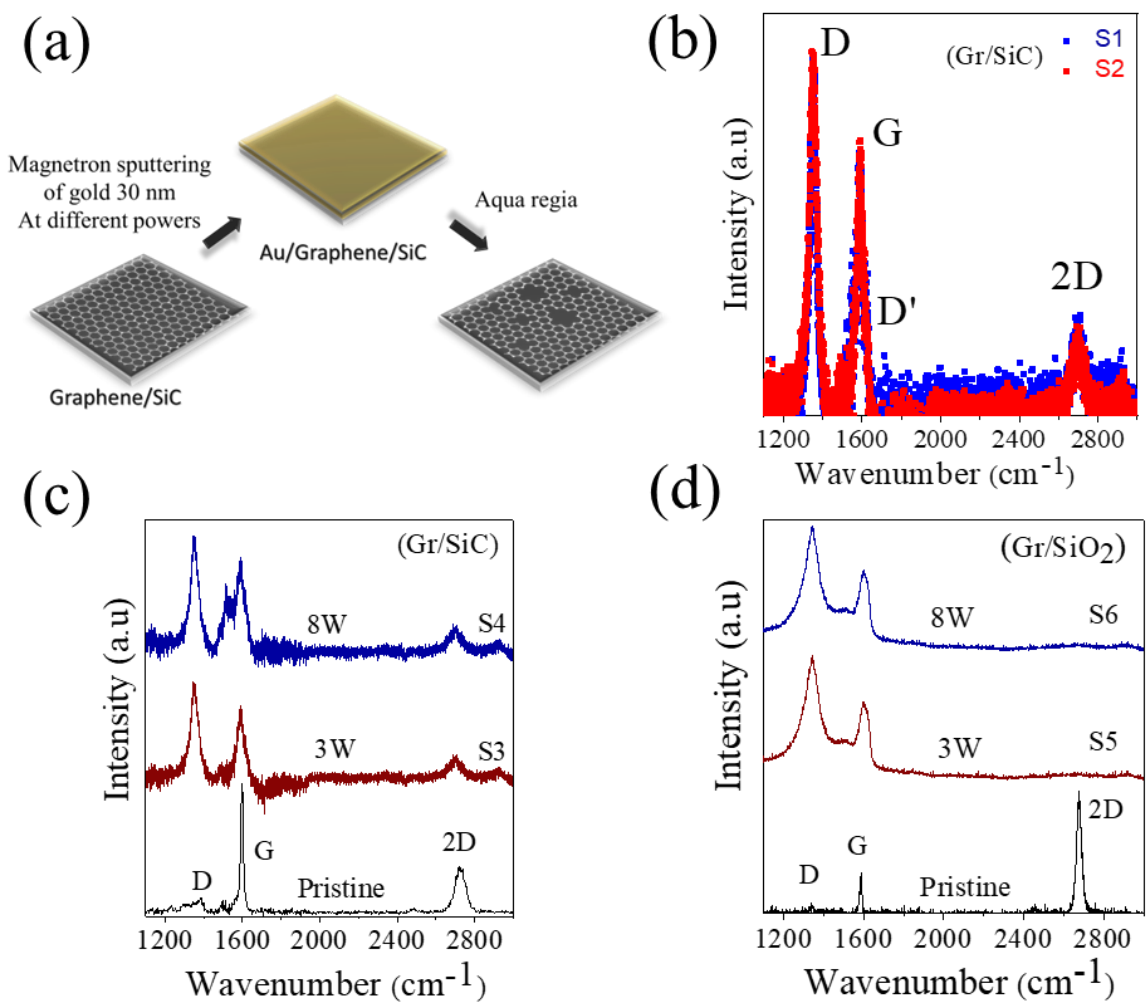

**Fig. 1 (a)** Schematic illustration of the sample preparation process **(b)** Measured Raman spectra of graphene on SiC of samples S$_1$ (blue) and S$_2$ (red). **(c)** Measured Raman spectra graphene on SiC of samples S$_3$ (blue), S$_4$ (red), and the as-grown graphene on SiC (black). **(d)** Measured Raman spectra of graphene on SiO$_2$ of samples S$_5$ (blue), S$_6$ (red), and the as-grown graphene on SiO$_2$ (black). All Raman measurements are taken at EL = 2.33 eV (λL = 532 nm).

Samples S$_1$, S$_2$, S$_3$, and S$_4$ correspond to graphene grown on SiC. At the same time, samples S$_5$ and S$_6$ are related to graphene grown on SiO$_2$. Figure 1-(b) shows the Raman spectra of two samples S$_1$ and S$_2$, sputtered by the same sputtering power under the same conditions. The samples clearly show the presence of a large defect peak (D) at 1350 cm$^{-1}$, higher than the peak of G. The same I$_D$/I$_G$ ratio (I$_D$/I$_G$=1.3) in the Raman lines of S$_1$ and S$_2$ indicates that they contain the same defect concentration. This reveals the possibility of efficiently controlling the defect density in graphene.

The samples S$_3$, S$_5$ (S$_4$, S$_6$) were treated by 3W (8W) sputtering powers, respectively. Figure 1-(c) shows the Raman spectra for samples S$_3$ (red), S$_4$ (blue), and the as-grown graphene (black). The spectra are represented after subtracting the SiC substrate Raman lines. The as-grown graphene shows a small D-band and a low I$_D$/I$_G$ ratio of about 0.1, which reveals high crystallinity. The sputtered samples show an intense D-band due to the defects generated during the sputtering process. The I$_D$/I$_G$ ratio equals 1.5 and 1.4 for S$_3$ and S$_4$, respectively. Also, the spectrum shows small bands around 1450 cm$^{-1}$ and 1510 cm$^{-1}$.

Figure 1-(d) depicts the Raman spectra for samples S$_5$ (red), S$_6$ (blue), and for as-grown graphene (black). The pristine graphene shows a small D-band, and the I$_D$/I$_G$ ratio is approximately 0.12. Due to the disorder caused by the generated defects during the sputtering process, samples S$_5$ and S$_6$ show an intense D-band. The I$_D$/I$_G$ ratio is respectively equal to 1.8 and 1.9. The Raman spectra of samples S$_5$ and S$_6$ clearly show the tiny band around 1450 cm$^{-1,}$ as observed for samples S$_3$ and S$_4$. Such band is already observed for defective graphene[11] and amorphous graphite thin films,[12-13,] and is often attributed to a third-order Raman mode of the silicon substrate. However, this band

is also observed for nano-porous graphene deposited on polyethylene terephthalate foil.[14] Hence, this could be due to pores in graphene-based samples since it was observed for amorphous, defective, and nanoporous graphene.

We calculated nonresonant Raman spectra of nanoporous monolayer graphene to estimate the vacancy size generated during the sputtering. The pores have been randomly distributed with a condition of a minimum distance ($d_{min} > 30$ Å) to be considered between every two pores to avoid their interactions. The simulation starts with a single pore and then considers randomly distributed nanopores to approach the measured Raman spectra. The best nanopores configuration fitting the measured Raman spectra are about 7.5 Å to 12.5 Å, illustrated in Figure 2 (b).

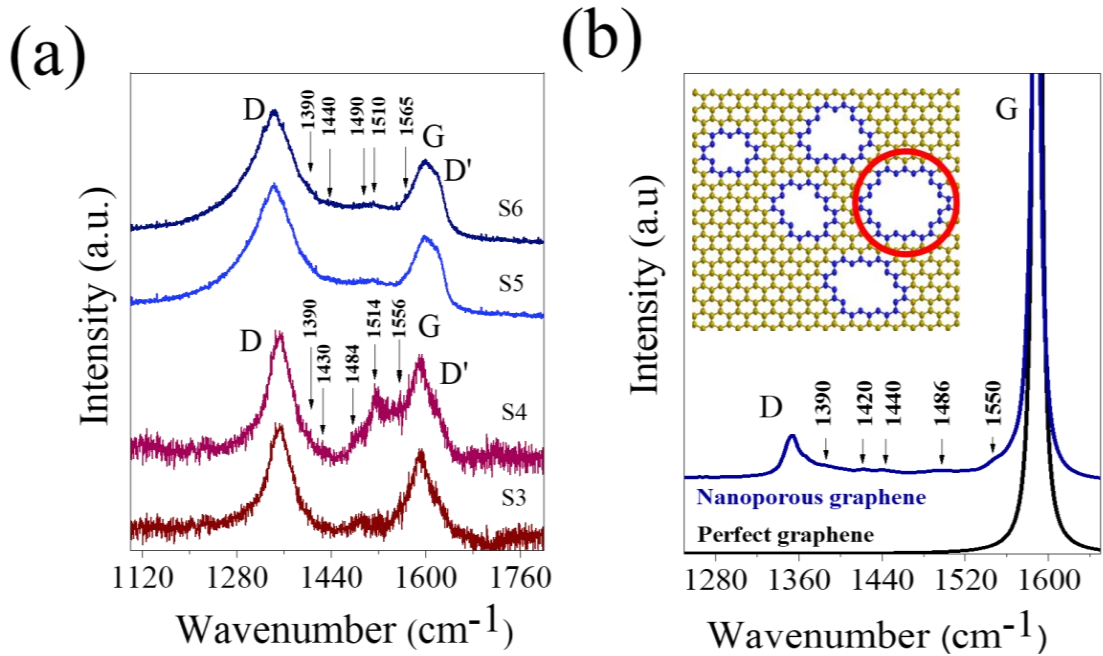

**Fig.2 (a)-** Measured Raman spectra of samples S3, S4, S5, and S6 in the frequency range of 1100cm$^{-1}$ to 1800cm$^{-1}$. **(b)**- Calculated Raman spectra of perfect graphene and graphene with nanopores. Inset: Schematic structure of graphene with nanopores. All Raman measurements are taken at EL = 2.33 eV (λL = 532 nm).

Figure 2-(a) and (b) represent the measured and calculated spectra for samples $S_3$, $S_4$, $S_5$, and $S_6$ for perfect and defective graphene in 1100 cm$^{-1}$ to 1800 cm$^{-1}$ frequency range. It is observed that the measured and calculated spectra show a high agreement regarding the positions of the D and G bands. We should note that the computed spectra show a small band around 1390 cm$^{-1}$, 1450 cm$^{-1}$, 1486 cm$^{-1}$, and 1510 cm$^{-1}$. We found that when the distance between defects becomes smaller, these small bands form a single large band around 1450 cm$^{-1}$, observed in the spectra of samples $S_5$ and $S_6$. This confirms that these small bands are due to the graphene with pores rather than the substrate effects, as reported by different research groups.[11-12]

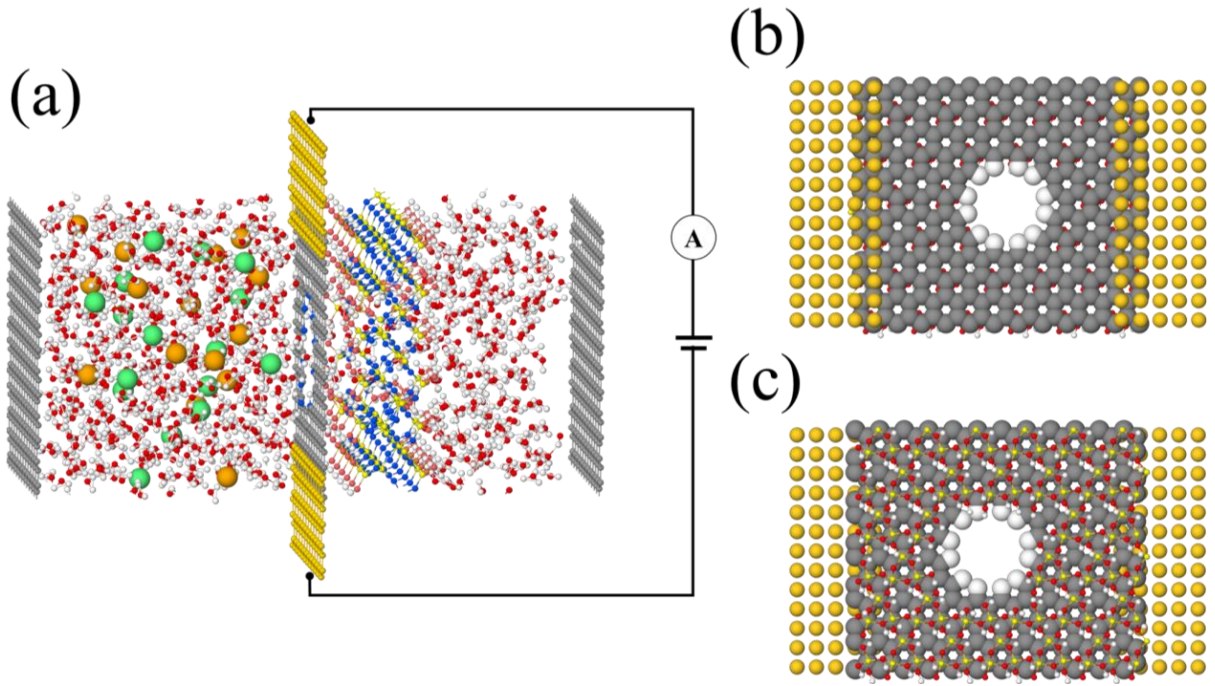

**Fig. 3: (a)** Schematic of the proposed graphene device-based water filtration membrane (The yellow atoms representing the electrodes are added to depict the presence of the applied electric field.), (b) Top and (c) bottom view of the 2E-NPG/SiO2 membrane

Figure. 3 shows the schematic concept model of the filtration with the applied electric field parallel to the membrane (Perpendicular direction to the water flow). The 2E-NPG/SiO$_2$ or 2E-NPG/SiC membrane is in the center of the simulation box. The right section contains an aqueous solution of water and NaCl with a salt concentration of 72 g/L, twice as high as seawater (~35 g/L), to achieve more significant ion–pore interactions. The edges of the simulation box contain graphene pistons at which pressure is applied. The substrate has the same nanopore morphology as graphene. The pore diameter is about ~0.82 nm, with the geometry highlighted in Figure 2 (b) inset.

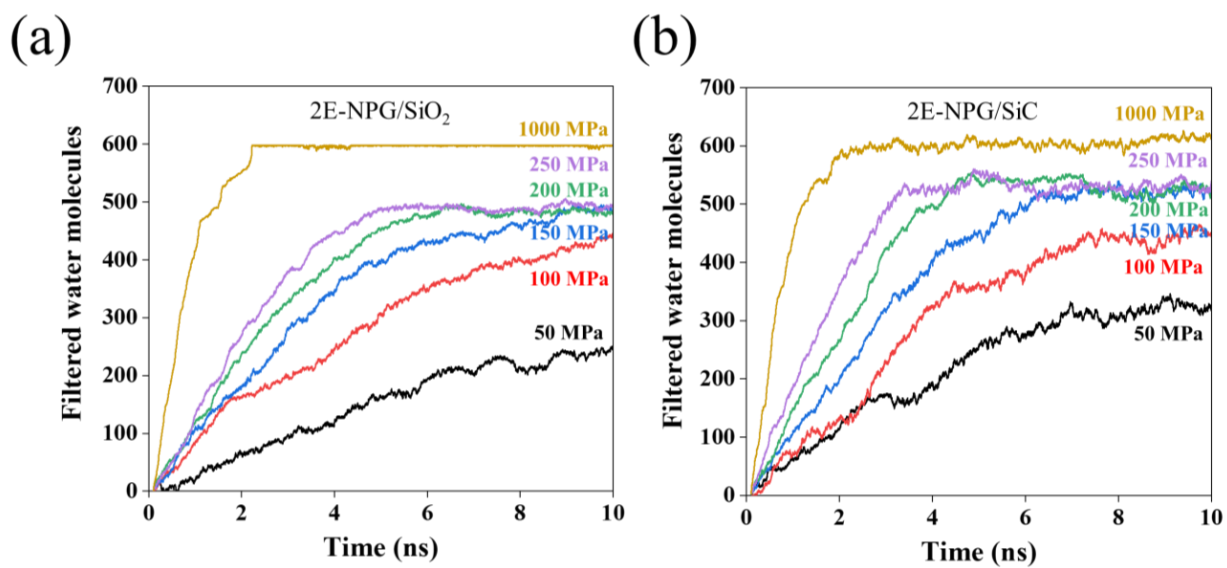

**Fig. 4 (a)** Number of filtered water molecules as a function of the simulation time for the 2E-NPG/SiO$_2$ at different applied pressure **(b)** Number of filtered water molecules as a function of the simulation time for the 2E-NPG/SiC at different applied pressure.

Figure 4 shows the water flow rate through the 2E-NPG/SiO$_2$ and 2E-NPG/SiC as a function of time at different applied pressures from 50 MPa to 1 GPa with no applied electric field. The flow rate is constant over time and increases with the applied pressure; this acts as a driving force for molecules' passage. Each curve starts with a linear regime in which water molecules pass at a constant rate; then, each trajectory reaches a saturation point due to the feed side being depleted before the simulation's end for specific pressure values. Having a linear pattern means the effects of the finite size of the simulation box and the higher feed salinity due to water being filtered through the membrane are insignificant. The slope of each flow curve indicates the water flow rate over a certain period and is proportional to the applied pressure. We note that 2E-NPG/SiC allows more water molecules to pass than 2E-NPG/SiO$_2$. The permeability parameter was extracted from the slope of the water flux as a function of pressure (Figure S1 in Supplementary Information) and is reported in Table 1. The time intervals are 0 to 3 ns to remain in the transitory regime. The approximate permeability value for 2E-NPG/SiO$_2$ is about 21.46 L/day.cm².MPa, which corresponds to 894.16 L/h.m².bar and 29.00 L/day.cm².MPa for 2E-NPG/SiC, which is equivalent to 1208 L/h.m².bar. For the substrate-free 2E-NPG, the permeability obtained is 2015 L/h.m².bar.

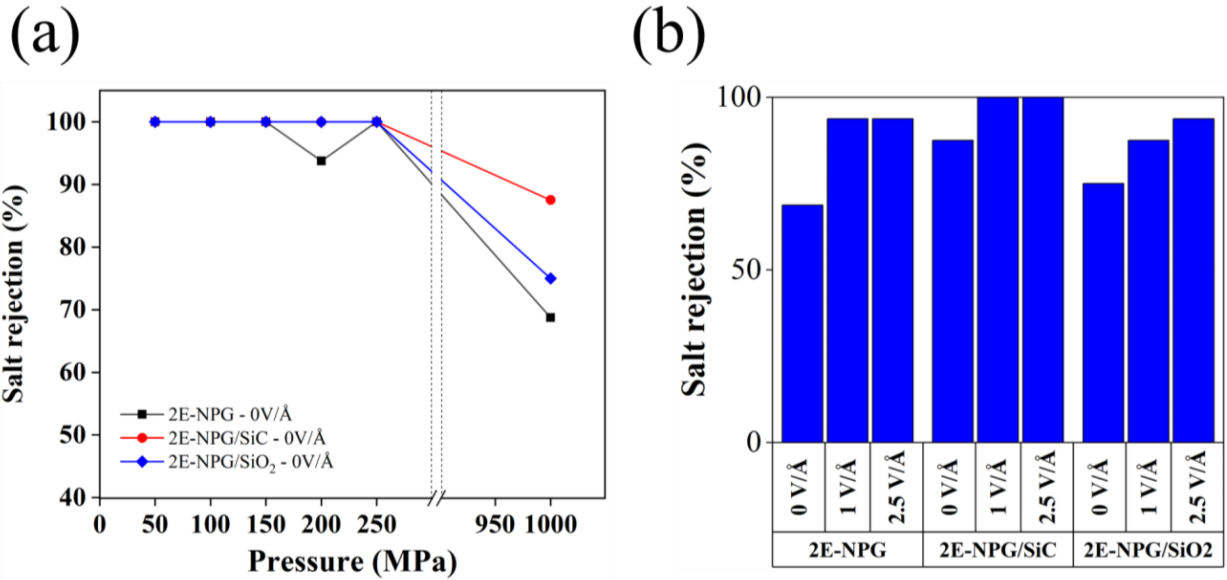

**Fig. 5 (a)** Salt rejection as a function of pressure with no electric field **(b)** Salt rejection as a function of applied electric fields for 1000 MPa.

The permeability does not depend on the applied electric field since the water molecule flux is not dependent on the perpendicular electric field up to 2.5V/Å. Therefore, the pores should be larger than a specific critical size to allow the permeation of water molecules inside the graphene and the substrates. However, to effectively prevent the passage of salt ions, their diameter must not exceed a given limit, and we can increase this limit by applying an electric field (perpendicular to water flux).

To assess the passage of NaCl ions inside the nanopore, we calculated the salt rejection rate $Sr$ from the following equation at different electric field:[15]

$$Sr = \left(1 - \frac{Np}{Nf}\right) \times 100$$

$N_p$ and $N_f$ are the percentage of salt in the permeate and feed region at $t_{1/2}$, corresponding to when half of the water molecules have flowed to the permeate side at each given pressure (transitory regime).

Figure 5 (a) shows the salt rejection rate without an applied electric field. From 50 to 1000 GPa, the 2E-NPG/SiC has a salt rejection of around 100 % at low pressure, higher than in previous studies[4]. A graphene membrane free-substrate (2E-NPG) was also simulated for comparison, and we found that the substrate enhances salt rejection. At higher applied pressure (>250 MPa), the salt rejection decreases but is still very high for 2E-NPG/SiC as compared to 2E-NPG/SiO$_2$ or other reported NPG membranes.[16]

Figure 5 (b) shows the variation of salt rejection for applied electric fields from 1 to 2.5V/Å for a pressure of 1000 MPa. As observed, an electric field induces ions separation, reduces ions' passage through the pore[15,17], and enhances the salt rejection rate. In addition, it was reported that the permeability value increased twice by increasing the nanopore diameter from 12.6 Å to 19.08 Å.[18.] Our concept makes it possible to have the highest permeability value with keeping a 100 % salt rejection.

To better understand the water molecule filtration near the membrane and substrate pores, we calculated the 2D average density of oxygen atoms. The water molecules distribution near the pores dominates water permeability and salt rejection. The water structure is determined by pore size effects, substrate, electric field, and chemical effects by pore functionalization.

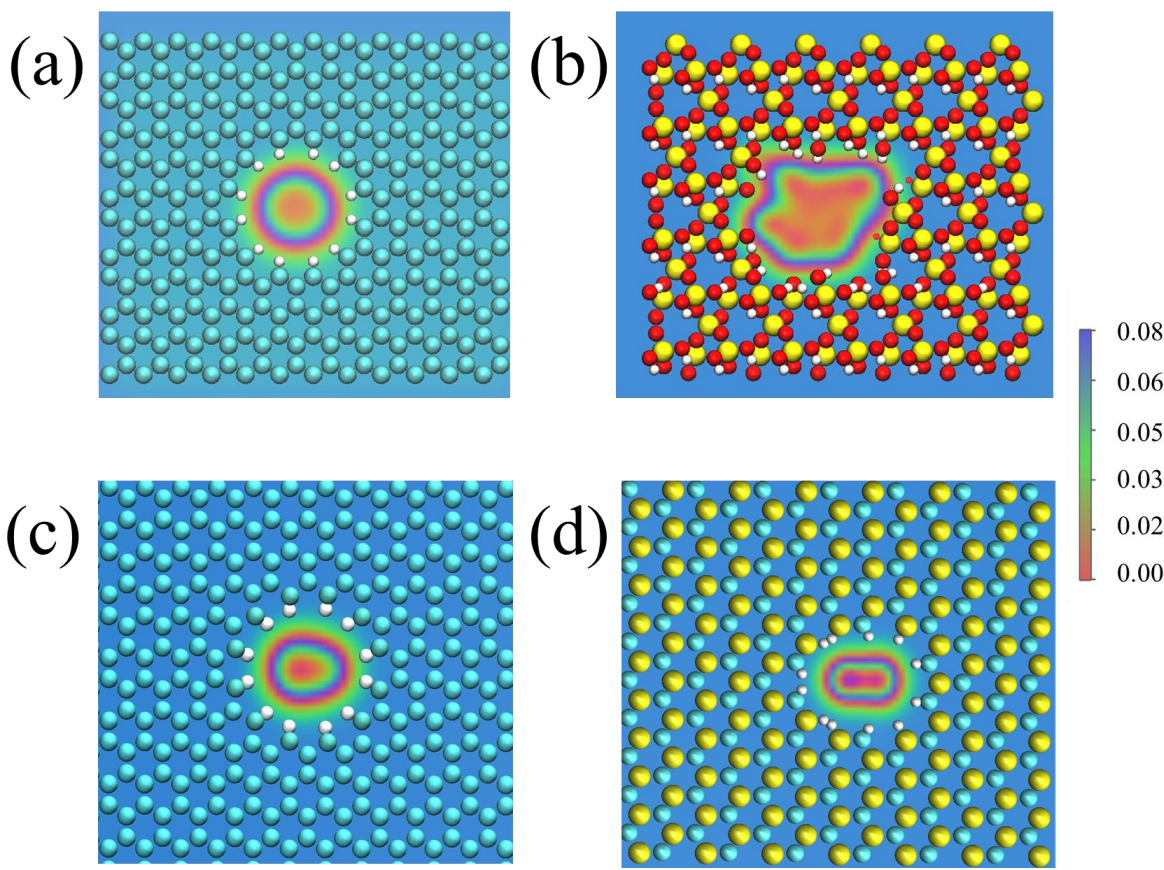

**Fig. 6 (a)** Oxygen density map at the vicinity of the graphene pore for the 2E-NPG/SiO$_2$ without electric field **(b)** Oxygen density map at the vicinity of the substrate pore for the 2E-NPG/SiO$_2$ without electric field **(c)** Oxygen density map at the vicinity of the graphene pore for the 2E-NPG/SiC without electric field **(d)** Oxygen density map at the vicinity of the substrate pore for the 2E-NPG/SiC without an electric field. Red indicates the region where no water oxygen is found, while blue regions indicate the highest probability of finding an oxygen atom.

Figure 6 shows the oxygen density map at the graphene and substrate pores. The hydrophobic nature of the hydrogenated group makes the oxygen trajectory close to the pore edges. The same behavior can be observed for the substrate pore. The observed densities are comparable in the graphene regions for both substrates. However, these densities are different at the substrate. The oxygen density region for the $SiO_2$ is more significant than in SiC; Still, the calculated permeabilities show a better performance for the 2E-NPG/SiC. This physical picture suggests we must also consider the activation energy due to the substrate effect since the energy barriers govern the water transport through a nanopore, the energy that water molecules must overcome[19].

Overall, this work answers two of the main challenges of graphene-based desalination membrane, its mechanical stability under applied high pressure and its pore size distribution easily achieved here by sputtering, where SiC or $SiO_2$ substrate supports the thin-film active layer. Indeed, this study shows that graphene monolayers on substrates with an electric field can withstand pressures upward of 1000 MPa without ripping, with a high permeability and salt rejection value. Another advantage of the electric field is that it can be used to evacuate and clean salt deposition around the nanopore.

The second remaining challenge is physically creating an extremely narrow pore size distribution in the SiC and the $SiO_2$ substrate.

In summary, we propose and indicate that 2E-NPG/SiC and 2E-NPG/$SiO_2$ concepts with an applied electric field can reject salt ions while letting water flow at permeabilities several orders of magnitude higher than existing RO membranes and are more realistic, efficient, robust, and mechanically stable at high pressure than other graphene-based concepts. We show that we can create and control the nano-pore size using sputtering. We found that the Raman mode (around 1450 cm$^{-1}$) was reported and attributed in the literature to the substrate effect due to graphene defects. This work highlights the promise of graphene on a substrate for water desalination. Our approach strongly suggests investigating low dimensional Materials in general for desalination membranes can yield significant improvements over existing membranes. This work will add substantial inputs to the next generation of membranes for clean water technology.

**Methods**

**Raman simulation**

The main parameters for the Raman spectra simulation are the dynamic matrix (D) and the polarizability tensor of the molecular system, which allow the determination of the Raman active modes' positions and their intensities. D describes the interactions between all atoms in the system; it has a 3N × 3N dimension, where *N* is the number of atoms in the molecule. In practice, the system's symmetry can reduce the number of matrix variables to be calculated. In our case, the random distribution of the pores causes a low symmetrical aspect of the system. Hence only the Hermitian character of D will be exploited. This requires computing all the elements of D (diagonal and off-diagonal), which increases the computational time. Under these constraints associated with the critical number of atoms (in our case, 20000 atoms), the diagonalization of the matrix becomes either impossible or requires a long computational time. However, the elements of our system's dynamic matrix are obtained using the constant force model introduced by Wirtz and Rubio,[20] which describes the interactions between the carbon atoms located away from the pores. While the interactions between the carbon atoms of the vacancies edges are calculated using density functional theory (DFT) as implemented in the SIESTA package.[212]

The Raman response is intimately related to molecular polarizability fluctuations. We used the bond polarizability model BPM to calculate the Raman lines intensities.[22] This model is widely used to calculate the Raman spectra for different systems, and it shows perfect accuracy with the experimental results. The simulation of our Raman spectra is based on the spectral moment method (SMM). This one allows us to calculate the Raman spectra without D diagonalization. The SMM, Raman equations and computational details to derive the Raman susceptibility from the BPM are included in these references.[23-25]

**Molecular dynamics simulations**

All the simulations were performed with the LAMMPS package.[26] First, the porous 2E-NPG/$SiO_2$ and 2E-NPG/SiC membranes were created. To do that, a careful scanning of the lattice parameter was done to ensure minimal mismatch between graphene and substrate to respect the periodic boundary conditions. The system consisted of a finite box measuring 9 nm along the z direction and ~2.8, ~3.25 nm along the x and y directions. The graphene membrane was fixed in the middle of the box at 4.5 nm. Six hundred water molecules were filtered with 16 Na and 16 Cl ions, corresponding to a salt concentration of 72 g/L. Two pistons represented by graphene sheets were placed on the box's top and bottom sides, with the piston's main role allowed to push water toward the membrane at different external pressures. Pressures from 50 MPa up to 1000 MPa were applied to the piston.

Before running dynamics on our system, a minimization of the box size, followed by a relaxation of water molecules, was allowed by using an NVT ensemble.

For the dynamics part, an NVT ensemble at a temperature of 300 K was used with a timestep of 1 fs which was deemed sufficient for reproducible results with good convergence. The total simulation time is up to 10 ns. To represent all the interactions within our system, three potentials were used. Tersoff potential for Si/O/H was used, AIREBO potential was used for the carbon atoms and the functionalized hydrogen group, and Lennard Jones TIP4P for water molecules and interatomic interactions. It should also be mentioned that the shape and size of the pore created in the membrane were chosen based on the experimental and Raman results, which suggest a circular form of our pores with a length of 1 nm. We used this simulation technique in our previous work on phosphorene numerical synthesis[27].

**Acknowledgments:** OCP Foundation has supported this work with the project grant AS70, "Towards phosphorene-based materials and devices, and with the support of the Chair "Multiphysics and HPC" led by Mohammed VI Polytechnic University. We acknowledge the High-Performance Computing (HPC) Facility of Mohammed VI Polytechnic University – Toubkal.

**Authors' contributions:** A.E.F. designed the research concepts from the experimental approach to the simulation part. S.A.A.A carried out the Raman simulation. I.B. and M.A. carried out molecular dynamics simulations. All authors discussed the results and wrote the paper.

**Ethics declarations:** The authors declare no competing interests.